\documentclass[fleqn,10pt]{wlscirep}
\usepackage[utf8]{inputenc}
\usepackage[T1]{fontenc}
\usepackage{amsmath}
\usepackage{amsfonts}
\usepackage{amssymb}
\usepackage{amsthm}

\newtheorem{theorem}{Theorem} 
\newtheorem{proposition}[theorem]{Proposition}

\title{Homomorphic Polynomial Public Key Cryptography for Quantum-secure Digital Signature}

\author[1,*]{Randy Kuang}
\author[1]{Maria Perepechaenko}
\author[2]{Mahmoud Sayed}
\author[1]{Dafu Lou}
\affil[1]{Quantropi Inc., Ottawa, K1Z 8P9, Canada}
\affil[2]{Systems and Computer Engineering, Carleton University, Ottawa, K1S 5B6, Canada}

\affil[*]{randy.kuang@quantropi.com}

\begin{abstract}
In their 2022 study, Kuang et al. introduced the Multivariable Polynomial Public Key (MPPK) cryptography, a quantum-safe public key cryptosystem leveraging the mutual inversion relationship between multiplication and division. MPPK employs multiplication for key pair construction and division for decryption, generating public multivariate polynomials. Kuang and Perepechaenko expanded the cryptosystem into the Homomorphic Polynomial Public Key (HPPK), transforming product polynomials over large hidden rings using homomorphic encryption through modular multiplications. Initially designed for key encapsulation mechanism (KEM), HPPK ensures security through homomorphic encryption of public polynomials over concealed rings. This paper extends its application to a digital signature scheme. The framework of HPPK KEM can not be directly applied to the digital signatures dues to the different nature of verification procedure compared to decryption procedure. Thus, in order to use the core ideas of the HPPK KEM scheme in the framework of digital signatures, the authors introduce an extension of the Barrett reduction algorithm. This extension transforms modular multiplications over hidden rings into divisions in the verification equation, conducted over a prime field. The extended algorithm non-linearly embeds the signature into public polynomial coefficients, employing the floor function of big integer divisions. This innovative approach overcomes vulnerabilities associated with linear relationships of earlier MPPK DS schemes. The security analysis reveals exponential complexity for both private key recovery and forged signature attacks, taking into account that the bit length of the rings is twice that of the prime field size. The effectiveness of the proposed Homomorphic Polynomial Public Key Digital Signature (HPPK DS) scheme is illustrated through a practical toy example, showcasing its intricate functionality and enhanced security features.
\end{abstract}
\begin{document}

\flushbottom
\maketitle
%
%
\thispagestyle{empty}


\section{Introduction}

Kuang in 2021~\cite{kuang2021ACCC} proposed a public key scheme called deterministic polynomial public key or DPPK for key exchange. DPPK essentially consists of two solvable univariate polynomials  $f(x)$ and $h(x)$ multiplying with a randomly chosen polynomial $B(x)$. The public key is constructed through polynomial multiplications over a finite field $\mathbb{F}_p$ excluding their constant terms. It is secure against the key recovery from the public key but it is insecure against the secret recovery with the deterministic factoring technique reported by Evdokimov in 1994 using Generalized Riemann Hypothesis or GRH~\cite{evdokimov1994factorization}. Later, Kuang and Barbeau~\cite{kuang2021performance,kuang2021CCECE} proposed to change the univariate polynomial $B(x)$ to a multivariate polynomial $B(x, u_1, \dots, u_m)$ with variable $x$ for the secret and $u_1, \dots, u_m$ for random noise values to produce two multivariate polynomials $P(x, u_1, \dots, u_m)$ and $Q(x, u_1, \dots, u_m)$ by excluding both constant terms and highest order terms. This updated variant is called the multivariate polynomial public key or MPPK for key encapsulation mechanism or KEM. In MPPK KEM, the excluded terms in polynomials $P(x, u_1, \dots, u_m)$ and $Q(x, u_1, \dots, u_m)$ form separate public key elements called noise functions, namely $N_0(u_1, \dots, u_m)$ associated with the constant terms and $N_n(x, u_1, \dots, u_m)$ associated with the highest order terms. Both $N_0(u_1, \dots, u_m)$ and $N_n(x, u_1, \dots, u_m)$ are encrypted with randomly chosen private key elements $R_0$ and $R_n \in \mathbb{F}_p$. By leveraging the noise functions together with all public key elements, the Gaussian elimination technique can be used to eliminate unknown coefficients from $B(x, u_1, \dots, u_m)$ and create equation system of unknown coefficients from univariate polynomials $f(x)$ and $h(x)$, potentially reducing the private key security. 

Kuang, Perepechaenko, and Barbeau in 2022 ~\cite{kuang2022-MPPK-KEM} proposed to encrypt the coefficients of the noise functions $N_0(u_1, \dots, u_m)$ and $N_n(x, u_1, \dots, u_m)$ over a hidden ring $\mathbb{Z}/S\mathbb{Z}$ with self-shared secret key values $R_0$ and $R_n$ respectively. The ring size is required to be at least $2\times$ bigger than the size of the field prime $p$. Kuang and Perepechaenko in 2023~\cite{hppk-f1000-2023} further extended this MPPK KEM by encrypting the entire public key elements over a hidden ring and renamed it as Homomorphic Polynomial Public Key or HPPK KEM because the encryption over the hidden ring holds the homomorphic property. Kuang and Perepechaenko in 2023~\cite{kuang2023-HPPK-KEM} proposed a further variant of the HPPK KEM over dual hidden rings with each public polynomials $P(x, u_1, \dots, u_m)$ and $Q(x, u_1, \dots, u_m)$ encrypted over their own separate rings. The homomorphic symmetric encryption essentially turns public key cryptography into symmetric encryption with the self-shared keys.

On the other hand, Kuang, Perpechaenko, and Barbeau in 2022~\cite{kuang2022.08.01-DS} proposed their digital signature scheme or MPPK/DS based on the identity equation $f(x)Q(x, u_1, \dots, u_m) = h(x)P(x, u_1, \dots, u_m) \bmod{p}$, with $P(x, u_1, \dots, u_m)=f(x)B(x, u_1, \dots, u_m)$ and $Q(x, u_1, \dots, u_m)=h(x)B(x, u_1, \dots, u_m)$ over the prime field $\mathbb{F}_p$. 
Later, Kuang and  Perpechaenko optimized MPPK/DS for parameter selections~\cite{kuang2023-ODS}. The identity equation in MPPK/DS offers a way to establish the signature verification but it leaves a way for the forged signature reported later by Guo in 2023~\cite{Guo2023}. 

\section{Contribution}
In this paper, we propose a novel digital signature scheme based on the HPPK KEM using the symmetric encryption over dual hidden rings with secret keys $R_1$ from $\mathbb{Z}/S_1\mathbb{Z}$ and $R_2$ from $\mathbb{Z}/S_2\mathbb{Z}$ and considering 
\begin{align}\label{eq:identity1}
   R_1^{-1} R_1 = 1 \bmod{S_1}, R_2^{-1} R_2 = 1 \bmod{S_2}
\end{align}
with condition $gcd(R_1, S_1)=1$ and $gcd(R_2, S_2)=1$. We can then insert Eq.~\eqref{eq:identity1} into the identity equation $f(x)Q(x, u_1, \dots, u_m) = h(x)P(x, u_1, \dots, u_m) \bmod{p}$ and turn it into a conditional identity equation with knowing the secret keys $R_1 \in \mathbb{Z}/S_1\mathbb{Z}$ and $R_2 \in \mathbb{Z}/S_2\mathbb{Z}$. The secret keys $R_1, S_1$ and $R_2, S_2$ are used to encrypt coefficients of the product polynomials $P(x, u_1, \dots, u_m)$ and $Q(x, u_1, \dots, u_m)$, respectively, using modular multiplicative operations and the produced ciphertext coefficients are used to derive public key polynomials. 
The moduli $S_1$ and $S_2$ representing the hidden rings can't be shared with the signature verifier. We enhance the Barrett reduction algorithm by adapting it to convert modular multiplications into divisions, utilizing the Barrett parameter $R=2^k$, where $k$ significantly surpasses the bit length of $S_1$ and $S_2$. This adaptation addresses the potential issue wherein the floor function of the Barrett reduction algorithm may occasionally yield a result $z-1$ instead of $z$ due to the limitations of the floor function's cutoff.  Then the conditional identity equation with $mod \ p$ could be partially applied to the moduli $S_1$ and $S_2$ to produce public key elements $s_1=\beta * S_1 \bmod{p}$ and $s_2=\beta * S_2 \bmod{p}$ with a random chosen $\beta\in \mathbb{F}_p$. The signature elements from the private polynomials $f(x)$ and $h(x)$ are calculated with $x=Hash(M)$ and then decrypted into $F$ and $H$ as signature elements with $R_2$  and $R_1$ respectively. 

Related work is reviewed in Section~\ref{sec:relwork}.
The HPPK-THR KEM cryptography system is briefly described in Section~\ref{sec:hppk}. Section~\ref{sec:hppk-DS} introduces HPPK DS. Section~\ref{sec:security} discusses the security analysis. 
We conclude with Section ~\ref{conclusion}.

\section{Related Work}\label{sec:relwork}
This work is closely related to the multivariate Merkle-Hellman knapsack cryptosystem~\cite{M-H-knapsack-1978}, introduced in 1978. The Merkle-Hellman scheme is characterized by a private superincreasing sequence $b_1, b_2, \dots, b_n$ and a binary vector space $\mathbb{F}_2^n$. The public key is generated using a coprime pair $M, W$ through modular multiplication, resulting in $a_i = W*b_i \bmod{M}$. Consequently, the public multivariate polynomial is expressed as $p(x_1, \dots, x_n)=\sum_{i=1}^n a_i x_i$, where $x_i \in [0, 1]$. Encryption involves mapping an n-bit secret message $s$ into a binary string $x_1, x_2, \dots, x_n$, evaluating the polynomial value $C=p(x_1, \dots, x_n)$, and producing a ciphertext. Decryption comprises symmetric decryption $S= W^{-1} * C$, followed by decoding $S$ into the plaintext message $s$ bit-by-bit using the superincreasing private sequence.

Initially, the Merkle-Hellman knapsack cryptosystem was believed to be secure, being based on the knapsack problem, considered NP-complete. However, its vulnerability was demonstrated by Shamir in 1982~\cite{Shamir-1978-break-MHK}. This vulnerability primarily stems from its superincreasing sequence of the private key and the binary vector space. Nguyen and Stern illustrated successful attacks on both Merkle-Hellman knapsack cryptosystem and its variant Qu-Vanstone scheme,  using  orthogonal lattice-reduction techniques in 1997~\cite{phong-knapsack-1997}.

The Multivariate Public Key Cryptosystem (MPKC)~\cite{Ding2009mpkc} can be viewed as an extension of the Merkle-Hellman knapsack scheme. It replaces the binary vector space with a small finite field vector space $\mathbb{F}_q^n$, a single linear map with a bilinear map involving multiple multivariate polynomials, and eliminates the need for the superincreasing sequence. MPKC employs two reversible linear affine maps for symmetric encryption. By adopting these modifications, MPKC achieves security based on the Multivariate Quadratic (MQ) problem, a new NP-complete problem. However, this comes at the cost of larger public key and ciphertext sizes.

The HPPK Key Encapsulation Mechanism (KEM), proposed by Kuang et al. in 2023~\cite{kuang2023-HPPK-KEM}, can be seen as a fusion of the Merkle-Hellman knapsack system and MPKC. It retains the modular multiplication concept from the Merkle-Hellman scheme but operates over two hidden rings and replaces the single linear vector space over a small finite field with dual vector spaces over a larger finite field $\mathbb{F}_p$. One vector space is for polynomials representing messages, while the other is for linear vectors representing noises. These features enable HPPK cryptography to function as a KEM scheme~\cite{kuang2023-HPPK-KEM}. By eschewing the use of the superincreasing sequence and binary vector space for encoding and decoding, the attacks discovered in the Merkle-Hellman knapsack system are not directly applicable to HPPK cryptography. This paper further extend HPPK KEM for a digital signature scheme.

In the realm of Post-Quantum Cryptography (PQC), the field of digital signatures is currently undergoing extensive exploration and development. This is in response to the potential threats posed by quantum computers to traditional cryptographic algorithms. The National Institute of Standards and Technology (NIST) initiated the standardization process for Post-Quantum Cryptography (PQC) in 2017~\cite{NIST}, initially considering 69 candidates. The first round concluded in 2019, narrowing down the field to 26 candidates entering the second round~\cite{NIST8240}.

Among these candidates, only four Key Encapsulation Mechanism (KEM) options progressed to the third round: code-based Classic McEliece~\cite{McEliece1978}, lattice-based CRYSTALS-KYBER~\cite{KYBER}, NTRU~\cite{stehle2013making}, and SABER~\cite{SABER}. Additionally, three digital signature candidates advanced to the third round: lattice-based CRYSTALS-DILITHIUM~\cite{DILITHIUM} and FALCON~\cite{FALCON}, along with multivariate Rainbow~\cite{DingLUOV}, as shown in NIST status report in 2021~\cite{NIST8309}.
In a subsequent update, NIST declared Kyber as the sole KEM candidate for standardization~\cite{NIST8413}. For digital signatures, NIST identified CRYSTALS-Dilithium, FALCON, and SPHINCS+~\cite{SPHINCS} as the three standardized algorithms. 

In early 2022, vulnerabilities in the NIST round 3 finalists came to light. Damien Robert initiated this revelation by reporting an attack on Supersingular Isogeny Diffie–Hellman (SIDH) in polynomial time~\cite{sidh,sidh-broken1}. Subsequently, Castryck and Decru presented a more efficient key recovery attack on SIDH\cite{SIDH-broken}, achieving key recovery for NIST security level V in under 2 hours using a laptop.

In a separate development, a novel cryptoanalysis was introduced by Emily Wenger et al. in 2022~\cite{salsa-emily}. This approach utilizes Machine Learning (ML) for the secret recovery of lattice-based schemes. Wenger's team demonstrated that their attack could fully recover secrets for small-to-midsize Learning With Errors (LWE) instances with sparse binary secrets, up to lattice dimensions of $n=128$ with SALSA~\cite{salsa-emily}, $n=350$ with PICANTE~\cite{salsa-picante}, and $n=512$ with VERDE~\cite{salsa-verde}. The potential scalability of this attack to real-world LWE-based cryptosystems is being explored. The team is actively working on enhancing the attack's capability to target larger parameter sets, although the timeframe and resources required for this advancement remain uncertain. Their latest publishing not only improves  the secret recovery methods of SALSA but also introduce a novel cross-attention recovery mechanism~\cite{salsa-verde}. Their innovative attack has ushered in a new era of cryptoanalysis, particularly when integrating ML with quantum computing.

Sharp et al. recently achieved significant breakthroughs in integer factorization, as detailed in their report~\cite{memcpu-sharp2023}. Their research focuses on breaking a 300-bit RSA public key using the MemComputing technique in simulation mode. MemComputing, a novel computing paradigm characterized by time non-locality, represents a noteworthy departure from traditional computing methods. In contrast to quantum computing, which relies on physical quantum processing units and exponential quantum resources for substantial computational acceleration, MemComputing demonstrates the ability to achieve large prime factorizations with polynomial computing resources. This holds true even when using classical computing systems, although further acceleration can be attained through ASIC implementation. Notably, Zhang and Ventra have recently proposed a digital implementation of MemComputing~\cite{memcpu-zhang2023implementation}. The implications of this technological advancement suggest that classical cryptography may become vulnerable sooner than anticipated, potentially outpacing the threat posed by quantum computers.

\section{Brief of Homomorphic Polynomial Public Key Cryptography}\label{sec:hppk}

In this section, we are going to briefly describe the motivation behind and development of the HPPK scheme: the path from the original deterministic polynomial public key or DPPK, proposed by Kuang in 2021~\cite{kuang2021ACCC}, then multivariate polynomial public key or MPPK, proposed by Kuang, Perepechaenko, and Barbeau in 2022~\cite{kuang2022-MPPK-KEM}, to the homomorphic polynomial public key over a single hidden ring in 2023~\cite{hppk-f1000-2023}, and over a dual hidden ring scheme~\cite{kuang2023-HPPK-KEM}. MPPK schemes for digital signature or MPPK/DS have been proposed by Kuang, Perepechaenko, and Barbeau in 2022~\cite{kuang2022.08.01-DS}, then an optimized version was proposed by Kuang and Perepechaenko in 2023~\cite{kuang2023-ODS}. A forged signature attack was reported by Guo in 2023~\cite{Guo2023}. 
 
\subsection{DPPK}\label{DPPK}
The DPPK cryptography was first proposed by Kuang in 2021 to leverage the elementary identity equation~\cite{kuang2021ACCC} 
\begin{equation}
    \frac{f(x)}{h(x)} = \frac{B(x)f(x)}{B(x)h(x)} =\frac{P(x)}{Q(x)}\bmod{p}
\end{equation}
with $p$ being a prime, $f(x)$ and $h(x) \in \mathbb{F}_p$ as private solvable polynomials, and random base polynomial $B(x) \in \mathbb{F}_p$ used to encrypt $f(x)$ and $h(x)$. They are defined as follows over the field $\mathbb{F}_p$,
\begin{align}
    &f(x) = f_0 + f_1 x + \dots +f_\lambda x^\lambda,  \nonumber \\
    &h(x) = h_0 + h_1 x + \dots +h_\lambda x^\lambda, \\ 
    &B(x) = B_0 + B_1 x + \dots +B_n x^n.  \nonumber
\end{align}
Product polynomials $P(x)=f(x)B(x)$ and $Q(x)=h(x)B(x)$ are used to create public key polynomials $P'(x)= P(x)-P_0$ and $Q'(x)= Q(x)-Q_0$ by excluding their constant terms $P_0$ and $Q_0$ respectively. The randomly chosen secret $x\in\mathbb{F}_p$ is encrypted into ciphertext $\Bar{P}=P'(x)$ and $\Bar{Q}=Q'(x)$ and is extracted through a radical expression through following equation  
\begin{equation} \label{DPPK:dec}
    \frac{f(x)}{h(x)} = \frac{\bar{P}+f_0B_0}{\bar{Q}+h_0B_0} = k \ \text{mod p}\longrightarrow f(x)-kh(x)=0 \bmod{p}.
\end{equation}
Indeed, given that the decrypting party has knowledge of values $f_0B_0, h_0B_0, \bar{P}, \bar{Q}$, they can calculate value $k$, which can then be used in the equation
\begin{equation}
    \frac{f(x)}{h(x)} = k \mod p
\end{equation} to find $x$. 
Although the key recovery attack on the DPPK scheme is non-deterministic, and can be carried out by solving public key equation system using Gaussian elimination technique, DPPK is vulnerable to the secret recovery attack from the ciphertext under Generalized Riemann Hypothesis or GRH by Evdokimov~\cite{evdokimov1994factorization}.  

\subsection{MPPK Key Encapsulation Mechanism}
In order to overcome the DPPK vulnerability with respect to the secret recovery attack,  Kuang and Barbeau in 2021~\cite{kuang2021CCECE, kuang2021performance} 
proposed to replace the randomly chosen base polynomial $B(x)$ with a randomly chosen multivariate polynomial $B(x, u_1, \dots, u_m)$ over the prime field $\mathbb{F}_p$, called MPPK KEM. The variable $x$ denotes the secret and $u_1, \dots, u_m$ refer to the randomly chosen noise variables from $\mathbb{F}_p$. The public key polynomials $P'(x, u_1, \dots, u_m)$ and $Q'(x, u_1, \dots, u_m)$ are created from the product polynomials $P(x, u_1, \dots, u_m), Q(x, u_1, \dots, u_m)$ excluding their constant terms and highest order terms. The constant terms and the highest order terms are used to create two noise functions $N_0(u_1, \dots, u_m)=R_0B_0(u_1, \dots, u_m)$  and  $N_n(x, u_1, \dots, u_m)= R_n B_n(u_1, \dots, u_m)x^{n+\lambda} \in \mathbb{F}_p$, where $R_0$ and $R_n$ are the private key values, randomly selected from $\mathbb{F}_p$. 

Key encapsulation mechanism in MPPK is achieved through the randomly chosen noise variables $u_1, \dots, u_m \in \mathbb{F}_p$. The ciphertext consists of four elements $\Bar{P}=P'(x, u_1, \dots, u_m), \Bar{Q}=Q'(x, u_1, \dots, u_m), \Bar{N}_0=N_0(u_1, \dots, u_m)$ , and $\Bar{N}_n=N_n(x, u_1, \dots, u_m)$. And the decryption is the same as in DPPK
\begin{equation}\label{MPPK:k}
    k = \frac{f(x)}{h(x)} = \frac{B(x, x_1,\dots,x_m)f(x)}{B(x, x_1,\dots,x_m)h(x)} = \frac{P(x, x_1,\dots, x_m)}{Q(x, x_1,\dots,x_m)} \ \text{mod} \ p,
\end{equation}
where polynomial values of $P(x, u_1, \dots, u_m)$ and $Q(x, u_1, \dots, u_m)$ are calculated with four elements of ciphertext and private key: $f_0$, $h_0$, $R_0$, and $R_n$.

The two noise functions $N_0(u_1, \dots, u_m)$  and  $N_n(x, u_1, \dots, u_m) \in \mathbb{F}_p$ partially leak $B_{0j}$ and $B_{nj}$ because of $\frac{N_{0j}}{N_{0\ell}} = \frac{R_0B_{0j}}{R_0B_{0\ell}} =\frac{B_{0j}}{B_{0\ell}} \bmod{p}$ and $\frac{N_{nj}}{N_{n\ell}} = \frac{R_nB_{nj}}{R_nB_{n\ell}} =\frac{B_{nj}}{B_{n\ell}}$. Kuang, Perepechaenko, and Barbeau in 2022~\cite{kuang2022-MPPK-KEM} proposed a mechanism by choosing $R_0$ and $R_n$ to encrypt the coefficients of two noise functions over a hidden ring $\mathbb{Z}/S\mathbb{Z}$ with the bit length of the ring two times bigger than that of the prime field $\mathbb{F}_p$. Having the value $S$ hidden, implies that $S$ is a part of the private key. Without knowing $S$, the divisions $\frac{N_{0j}}{N_{0\ell}}$ and $\frac{N_{nj}}{N_{n\ell}}$ can not be carried out successfully. This eliminates the vulnerability originally present in the two noise functions.

The ciphertext of MPPK KEM with a hidden ring consist of elements but the elements $\Bar{N}_0$ and $\Bar{N}_n$ are three times bigger than the size of the prime field.

\subsection{HPPK KEM}
In the framework of the MPPK KEM scheme, the security of all coefficients within two noise functions was safeguarded through encryption over a hidden ring with a bit length $2\times$ larger than the prime field bit length. This concept was subsequently expanded to encompass all coefficients of product polynomials $P(x, u_1, \dots, u_m)$ and $Q(x, u_1, \dots, u_m)$ by Kuang and Perepechaenko in 2023~\cite{hppk-f1000-2023}. The encryption method has been demonstrated to be partially homomorphic, under addition and scalar multiplication operations. Consequently, the scheme was renamed as Homomorphic Polynomial Public Key, or HPPK.

In HPPK KEM, the entire product polynomials $P(x, u_1, \dots, u_m)$ and $Q(x, u_1, \dots, u_m)$ are considered private
\begin{equation}\label{eq:p-poly1}
\begin{aligned}
    &P(x, u_1, \dots, u_m) = f(x)B(x,  u_1, \dots, u_m) = \sum_{j=1}^{m} \sum_{i=0}^{n+\lambda} p_{ij} x^i u_j \bmod{p} \\
    &Q(x, u_1, \dots, u_m) = h(x)B(x,  u_1, \dots, u_m) = \sum_{j=1}^{m} \sum_{i=0}^{n+\lambda} q_{ij} x^i u_j \bmod{p}
\end{aligned}
\end{equation}
with $p_{ij}=\sum_{s+t=i}f_sB_{tj} \bmod{p}$ and $q_{ij}=\sum_{s+t=i}h_sB_{tj} \bmod{p}$. Choosing a secret $S$ with its bit length to be more than double of the prime field $\mathbb{F}_p$ and  secret encryption key $R_1$ and $R_2$ over the ring with condition $gcd(R_1, S) = 1$ and $gcd(R_2, S)=1$, we then encrypt entire coefficients of $P(.)$ and $Q(.)$ as follows
\begin{equation}
    p'_{ij} = R_1 \times p_{ij} \bmod{S}, q'_{ij} = R_2 \times q_{ij} \bmod{S}
\end{equation}
then the cipher polynomials can be written as
\begin{align}\label{eq:p-poly2}
    &\mathcal{P}(x, u_1, \dots, u_m) =  \sum_{j=1}^{m} \sum_{i=0}^{n+\lambda} p'_{ij}( x^i u_j \bmod{p})\nonumber \\
    &\mathcal{Q}(x, u_1, \dots, u_m) = \sum_{j=1}^{m} \sum_{i=0}^{n+\lambda} q'_{ij} ( x^i u_j \bmod{p})
\end{align}
The secret encryption is the same as in MPPK KEM, but now we only have two elements for the ciphertext: $\bar{\mathcal{P}}, \bar{\mathcal{Q}}$. The decryption is split into two phases: symmetric homomorphic decryption with $R_1$ and $R_2$
\begin{equation}
    \bar{P} = (R_1^{-1} \times \bar{\mathcal{P}} \bmod{S}) \bmod{p}, \bar{Q} = (R_2^{-1} \times \bar{\mathcal{Q}} \bmod{S}) \bmod{p}
\end{equation}
and then the secret extraction which is the same as in DPPK. In the most recent iteration of HPPK KEM, Kuang and Perepechaenko (2023) propose employing two distinct rings, $\mathbb{Z}/S_1\mathbb{Z}$ for the encryption of $P(.)$ and $\mathbb{Z}/S_2\mathbb{Z}$ for the encryption of $Q(.)$, to further enhance the security~\cite{kuang2023-HPPK-KEM}.

\subsection{MPPK DS}
A digital signature scheme based on MPPK has been proposed by Kuang, Perepechaenko, and Barbeau in 2022~\cite{kuang2022.08.01-DS} and later optimized variant was proposed by Kuang and Perepechaenko in 2023~\cite{kuang2023-ODS}. MPPK digital signature scheme originated from MPPK over a single prime field~\cite{kuang2021CCECE, kuang2021performance}. The signature verification equation stems from the identity equation
\begin{equation}\label{eq:ds-identity}
    f(x)Q(x, u_1, \dots, u_m) = h(x)P(x, u_1, \dots, u_m) \bmod{\varphi(p)},
\end{equation}
where $\varphi(p)$ is an Euler's totient function. For a prime $p$, $\varphi(p) = p-1$.
With the inclusion of two noise functions $N_0(u_1, \dots, u_m)=R_0B_0(u_1, \dots, u_m)$  and  $N_n(x, u_1, \dots, u_m)= R_n B_n(u_1, \dots, u_m)x^{n+\lambda} \in \mathbb{F}_p$, Eq.~\eqref{eq:ds-identity} becomes 
\begin{align}\label{eq:ds-identity1}
    f(x)Q'(x, u_1, \dots, u_m) =& h(x)P'(x, u_1, \dots, u_m) + s_0(x)N_0( u_1, \dots, u_m) \nonumber \\
                                &+ s_n(x)N_n(x,  u_1, \dots, u_m) \bmod{\varphi(p)}.
\end{align}
with $s_0(x)=f_0h(x)-h_0f(x), s_n(x)=f_{\lambda}h(x)- h_{\lambda}f(x) $. Choosing a secret base $g \in \mathbb{F}_p$ to avoid the discrete logarithm and performing modular exponentiation to the above Eq.~\eqref{eq:ds-identity1}, we obtain
\begin{equation}\label{eq:ds-identity2}
    A^{Q'(x, u_1, \dots, u_m)} = B^{P'(x, u_1, \dots, u_m)}C^{N_0( u_1, \dots, u_m)}D^{N_n(x,  u_1, \dots, u_m)} \bmod{p}
\end{equation}
with $A=g^{f(x)}, B=g^{h(x)}, C=g^{s_0(x)}, D=g^{s_n(x)}$ as signature elements. Recently, Guo uncovered a forged signature attack on MPPK DS~\cite{Guo2023}, attributed to the linearity between signature elements and public polynomials of Eq.~\eqref{eq:ds-identity1}. Attempts were made to address this vulnerability in MPPK DS, but it proved exceedingly challenging to rectify within a single modulo domain due to the linearity.

\section{HPPK Digital Signature Scheme}\label{sec:hppk-DS}
The homomorphic encryption operator $\hat{\mathcal{E}}_{(R, S)}$ was defined in~\cite{hppk-f1000-2023, kuang2023-HPPK-KEM} as
\begin{equation}\label{eq:he-en}
    \hat{\mathcal{E}}_{(R, S)}(f) = (R \circ f) \ \text{mod} \  S =f',
\end{equation}
with $f$ being a polynomial and decryption operator $\hat{\mathcal{E}}^{-1}_{(R, S)}$
\begin{equation}\label{eq:he-de}
    \hat{\mathcal{E}}^{-1}_{(R, S)}(f') = (R^{-1} \circ f') \ \text{mod} \  S.
\end{equation}
So it is very easy to verify that  
\begin{equation}\label{eq:he-de1}
    \hat{\mathcal{E}}^{-1}_{(R, S)}(f') = \hat{\mathcal{E}}^{-1}_{(R, S)}\hat{\mathcal{E}}_{(R, S)}(f)= (R^{-1} \times R \bmod{S}) \circ f = f.
\end{equation}
which reveals the unitary and reversible relation 
\begin{equation}\label{eq:unitary}
    \hat{\mathcal{E}}^{-1}_{(R, S)}\hat{\mathcal{E}}_{(R, S)} = 1.
\end{equation}
 Indeed, this encryption operator serves as a distinctive permutation operator or a quantum permutation gate that can be implemented in a native quantum computing system~\cite{kuang-epj-2022, Perepechaenko-epj-2023}. Furthermore, it adheres to the principles of non-commutativity, i.e. 
 \begin{equation}
    \hat{\mathcal{E}}_{(R, S)}\hat{\mathcal{E}}_{(R', S')} \neq \hat{\mathcal{E}}_{(R', S')}\hat{\mathcal{E}}_{(R, S)}.
\end{equation}
The non-commutativity is very important for us to reuse it multiple times to encrypt the entire coefficients of a product polynomial without reducing its security~\cite{qpp-springer-kuang-2022}, unlike the case of One-Time-Pad encryption with XOR because XOR operator is commutable.

\subsection{HPPK DS Algorithm}
Let's insert Eq.~\eqref{eq:unitary} into Eq.~\eqref{eq:ds-identity} with a randomly chosen $\alpha\in \mathbb{F}_p$
\begin{equation}\label{eq:ds-0}
    \{[\alpha f(x) \bmod{p}][\hat{\mathcal{E}}^{-1}_{(R_2, S_2)}\hat{\mathcal{E}}_{(R_2, S_2)}]Q(...)\} \bmod{p}= \{[\alpha h(x) \bmod{p}][\hat{\mathcal{E}}^{-1}_{(R_1, S_1)}\hat{\mathcal{E}}_{(R_1, S_1)}]P(...)\} \bmod{p} 
\end{equation}
then we reorganize Eq.~\eqref{eq:ds-0} into a verification equation
\begin{equation}\label{eq:verify}
    F(x) \mathcal{Q}(x, u_1, \dots, u_m) \bmod{p} =  H(x) \mathcal{P}(x, u_1, \dots, u_m) \bmod{p}
\end{equation}
with
\begin{align}
    &F(x) = \hat{\mathcal{E}}^{-1}_{(R_2, S_2)} [\alpha f(x) \bmod{p}] = R_2^{-1} \times [\alpha f(x) \bmod{p}] \bmod{S_2}   \\
    &H(x) = \hat{\mathcal{E}}^{-1}_{(R_1, S_1)} [\alpha h(x) \bmod{p}] = R_1^{-1} \times [\alpha h(x) \bmod{p}] \bmod{S_1}
\end{align}
as signature polynomials with $\alpha \in \mathbb{F}_p$ randomly chosen per signing message and
\begin{align}
    \mathcal{P}(x, u_1, \dots, u_m) &= \hat{\mathcal{E}}_{(R_1, S_1)} P(x, u_1, \dots, u_m) \nonumber \\
    &=\{\sum_{j=1}^m \sum_{i=0}^{n+\lambda} [R_1 \times p_{ij} \bmod{S_1}] (x^iu_j \bmod{p})\} \bmod{S_1}  \nonumber\\
    &=\{\sum_{j=1}^m \sum_{i=0}^{n+\lambda} P_{ij}  (x^iu_j \bmod{p})\}  \bmod{S_1}\\
    \mathcal{Q}(x, u_1, \dots, u_m) &= \hat{\mathcal{E}}_{(R_2, S_2)} Q(x, u_1, \dots, u_m) \nonumber\\
    &=\{\sum_{j=1}^m \sum_{i=0}^{n+\lambda} [R_2 \times q_{ij} \bmod{S_2}] (x^iu_j \bmod{p} ) \} \bmod{S_2} \nonumber \\
    &=\{\sum_{j=1}^m \sum_{i=0}^{n+\lambda} Q_{ij}  (x^iu_j \bmod{p}) \} \bmod{S_2}
\end{align}
as homomorphically encrypted polynomials with coefficients $P_{ij} = R_1*p_{ij} \bmod{S_1}$ and $Q_{ij} = R_2*q_{ij} \bmod{S_2}$. We can then rewrite the verification equation Eq.~\eqref{eq:verify} as 
\begin{align}\label{eq:verify1}
    \sum_{j=1}^m \sum_{i=0}^{n+\lambda} [F(x)Q_{ij} \bmod \ S_2] (x^iu_j \bmod{p}) \bmod{p} &= \sum_{j=1}^m \sum_{i=0}^{n+\lambda} [H(x)P_{ij} \bmod \ S_1] (x^iu_j \bmod{p}) \bmod{p}\nonumber \\
     \longrightarrow \sum_{j=1}^m \sum_{i=0}^{n+\lambda} V_{ij}(F) x^iu_j \bmod{p}  &= \sum_{j=1}^m \sum_{i=0}^{n+\lambda} U_{ij}(H) x^iu_j \bmod{p} \nonumber \\
     \longrightarrow V(F, x, u_1, ..., u_m) &= U(H, x, u_1, ..., u_m) \bmod{p}
\end{align}
with polynomial coefficients $V_{ij}(F)$ and $U_{ij}(H)$ to be defined as
\begin{equation} \label{eq:uv}
\begin{aligned}
    U_{ij}(H) &= [H(x)P_{ij} \bmod{S_1}] \bmod{p}  \\
    V_{ij}(F) &= [F(x)Q_{ij} \bmod{S_2}] \bmod{p}.
\end{aligned}
\end{equation}
The coefficients newly defined as $V_{ij}(F)$ and $U_{ij}(H)$ in Eq.\eqref{eq:uv} cannot be directly computed due to the hidden nature of moduli $S_1$ and $S_2$, as they are not accessible to the verifier. However, the Barrett reduction algorithm offers a solution, enabling the transformation of modular multiplications over both hidden rings into divisions\cite{barrett}:
\begin{equation}\label{eq:barrett}
    ab \bmod{n} = ab - n\lfloor \frac{ab}{n} \rfloor = ab - n\lfloor \frac{a \lfloor \frac{Rb}{n} \rfloor}{R} \rfloor = ab - n\lfloor \frac{a\mu}{R} \rfloor
\end{equation}
with 
 $\mu=\lfloor \frac{Rb}{n} \rfloor$ as the Barrett parameter and $R=2^k, k \geq \lceil \log_2 n \rceil$, the Barrett reduction algorithm proves instrumental in enhancing the efficiency of modular multiplications with the precomputed $\mu$. The outcome $z$ generated by the Barrett algorithm typically falls within the range of $[0, 2n)$ but not consistently within $[0, n)$. Therefore, if the result exceeds $n$, it necessitates returning $z - n$. Notably, it has been observed that the occurrence of $z>n$ cases can be mitigated by increasing the bit length of $R$. Our testing indicates that when $k- \lceil \log_2 n \rceil > 30$, the instances of $z>n$ tend to approach zero after 100,000,000 trials.

With Eq.~\eqref{eq:barrett}, we can transform Eq.~\eqref{eq:uv} into the following equations by multiplying a randomly chosen $\beta\in\mathbb{F}_p$ and then taking $\bmod{p}$ 
\begin{align}\label{eq:uv1}
    U_{ij}(H) &=  H(x)p'_{ij} -s_1\lfloor \frac{H(x)\mu_{ij}}{R} \rfloor \bmod{p}\nonumber \\
    V_{ij}(F) &= F(x)q'_{ij} -s_2\lfloor \frac{F(x)\nu_{ij}}{R} \rfloor \bmod{p}.
\end{align}
 with 
 
 \begin{equation}\label{eq:public}
 \begin{aligned}
     &s_1 = \beta S_1 \bmod{p}  \\
     &s_2 = \beta S_2 \bmod{p}  \\
     &p'_{ij} = \beta P_{ij} \bmod{p}  \\
     &q'_{ij} = \beta Q_{ij} \bmod{p}  \\
     &\mu_{ij} = \lfloor \frac{RP_{ij}}{S_1} \rfloor   \\
     &\nu_{ij} = \lfloor \frac{RQ_{ij}}{S_2} \rfloor
 \end{aligned}
 \end{equation}
to be elements of the \textbf{public key} for HPPK DS. The hidden $S_1, S_2$ are no longer required for the signature verification. The \textbf{private key} consists of:
\begin{equation}\label{eq:private}
\begin{aligned}
    &f(x), h(x): h_i, i=0, 1, ..., \lambda \\
    &R_1 \in \mathbb{Z}/S_1\mathbb{Z}    \\
    &R_2 \in \mathbb{Z}/S_2\mathbb{Z}    
\end{aligned}
\end{equation}
It should be pointed out that coefficients in Eq.~\eqref{eq:uv1} of public polynomials in Eq.~\eqref{eq:verify1}, especially the public key elements $\mu_{ij} = \lfloor \frac{RP_{ij}}{S_1} \rfloor$ and $\nu_{ij} = \lfloor \frac{RQ_{ij}}{S_2} \rfloor$ in Eq.~\eqref{eq:public} obtained from the Barrett reduction algorithm, exhibit non-linear associations with signature elements $F(x)$ and $H(x)$ at variable value $x$ (please note that $F(x)$ and $H(x)$ are considered as the numerical values of polynomial evaluations at message $x$), also referred to as signature-embedded coefficients. To elaborate further, the original verification equation in Eq.~\eqref{eq:verify} involves a specialized encryption via the Barrett reduction algorithm, which safeguards against attackers accessing symmetrically encrypted coefficients $P_{ij}$ and $Q_{ij}$. This inherent non-linearity plays a pivotal role in preventing the occurrence of forged signatures, as observed in the early stages of our MPPK DS scheme~\cite{Guo2023}. We intend to delve deeper into this aspect in the subsequent security analysis section, where we establish $\beta=1$ to ensure the independence of $s_1$ and $s_2$, along with selecting a randomly assigned value to govern the association between $s_1$ and $s_2$.

\subsection{Signing}
The signing algorithm is very straightforward in HPPK DS:
\begin{itemize}
    \item Assign the hash code of a message to variable $x=Hash(M)$;
    \item Randomly choose $\alpha \in \mathbb{F}_p$;
    \item Evaluate $F= R_2^{-1} \times [\alpha f(x) \bmod{p}] \bmod{S_2}$;
    \item Evaluate $H= R_1^{-1} \times [\alpha h(x) \bmod{p}] \bmod{S_1}$;
    \item HPPK DS $sig=\{F, H\}.$
\end{itemize}
Following this, the signer appends the signature $sig=\{F, H\}$ to the message, potentially including the public key if the recipient does not possess it. Upon receiving the message, the recipient can verify the signature using the public key of the HPPK DS. In scenarios where the hash code of the message exceeds the bit length of the finite field $\mathbb{F}_p$, segmentation becomes necessary. Each segment is aligned with $\mathbb{F}_p$, and a signature is independently generated for each segment. Subsequently, the segmented parts are concatenated to form the message's signature.

To prevent signature verification failure, it is advisable to conduct verification, as outlined in subsection~\ref{sec:verify}. In the event of verification failure, it is recommended to randomly select a new $\alpha \in \mathbb{F}_p$ and reevaluate $F$ and $H$. 

\subsection{Verify}\label{sec:verify}
The signature verification consists of two stages: evaluating signature embedded coefficients and comparing polynomial values at the variable value $x$. The procedure is as follows
\begin{itemize}
    \item $x=Hash(M)$, and randomly choose variables $u_1, \dots, u_m \in \mathbb{F}_p$;
    \item Evaluate $U_{ij}(H), V_{ij}(F)$ based on Eq.~\eqref{eq:uv1} for $i=0, ..., n+\lambda, j=1, ..., m$;
    \item Evaluate $V(F, x, u_1, ..., u_m)$ and $U(H, x, u_1, ..., u_m)$ and check if they are equal. If they are equal, the verification is passed. Otherwise, the verification is failed.
\end{itemize}

\subsection{A Variant of the Barrett Reduction Algorithm}
The Barrett reduction algorithm~\cite{barrett} is employed to enhance the efficiency of modular operations, particularly modular multiplications as expressed in Eq.~\eqref{eq:barrett}. The outcome $z$ typically lies within the range of $[0, 2n)$ due to the limitations imposed by the pre-computed floor function $\lfloor \frac{Rb}{n} \rfloor$ with $R=2^k$. In the context of the proposed HPPK DS scheme, however, it is essential for the result to fall within $[0, n)$. Notably, by increasing the value of $k$, the results obtained from the Barrett reduction algorithm can be effectively constrained to the desired range of $[0,n)$.

\begin{figure}[ht] 
\caption{A semi-log graph illustration of the Barrett reduction results falling in $[n, 2n)$ per $10^8$ computations of $z=a*b \bmod{n}$, with randomly chosen $a < n$ and $b < n$,   is plotted as a function of $\delta=k-\lceil log_2 n \rceil$ in bits. The blue line corresponds to $\lceil log_2 n \rceil= 208 $ bits, the yellow line to $\lceil log_2 n \rceil= 292 $ bits, and the grey line to $\lceil log_2 n \rceil= 400 $ bits. }
\includegraphics[scale = 0.50]{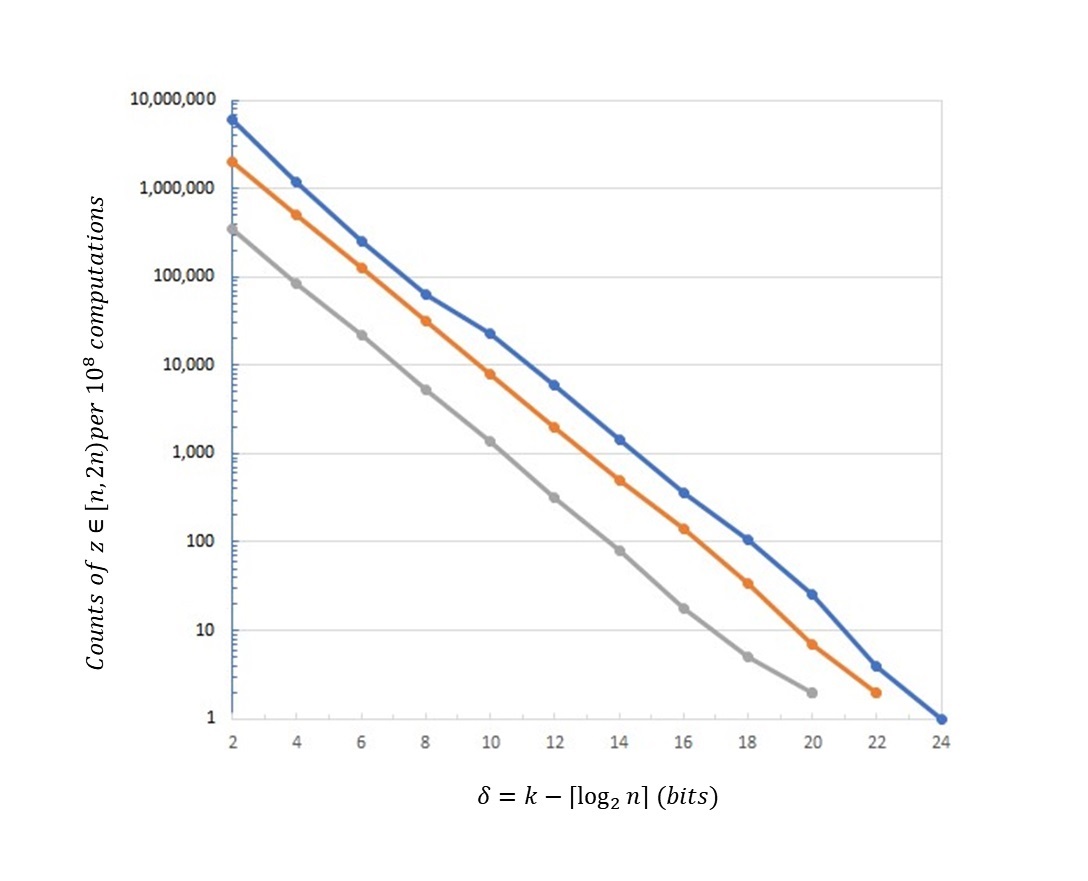}
\centering
\label{fig:barrett}
\end{figure}

All three lines exhibit a roughly linear relationship in the semi-log graph, converging to zero as $\delta$ surpasses 24 bits. However, as $k$ decreases to $\lceil \log_2 n \rceil$, the instances of falling beyond $[0,n)$ rise to $6\%$ for $\lceil \log_2 n \rceil=208$ bits, $2\%$ for $\lceil \log_2 n \rceil=292$ bits, and $0.3\%$ for $\lceil \log_2 n \rceil=400$ bits, respectively. Consequently, achieving a result within $[0,n)$ with high probability is feasible when $\delta=k-\lceil \log_2 n \rceil > 32$ bits.
In Fig.~\ref{fig:barrett}, a semi-logarithmic graph depicts the count of Barrett reduction results ($z=a*b \bmod{n}$) falling within the range $[n, 2n)$ per $10^8$ computations, where $a<n$ and $b<n$ are randomly chosen. The graph is plotted against the parameter $\delta=k-\lceil \log_2 n \rceil$. Three scenarios are represented by the blue line for $\lceil \log_2 n \rceil=208$ bits, the yellow line for $\lceil \log_2 n \rceil=304$ bits, and the grey line for $\lceil \log_2 n \rceil=400$ bits, respectively.

\subsection{A Toy Example}
 
\subsubsection{Key Pair Generation}
For purposes of simplicity, consider HPPK DS defined over a prime field $\mathbb{F}_{13}$ and choose $S_1=6797, S_2=7123$. we select $R_1=4267, R_2=6475$. Let the randomly chosen private key consist of the following values:
\begin{equation}\label{toy-sk}
    \begin{cases}
    S_1 = 6797, R_1 = 4267 \\
    S_2 = 7123, R_2 = 6475\\
    f(x) = 4+9 x\\
    h(x) = 10+7 x\\
    \end{cases}
\end{equation}
Let the base polynomial $B(.)$ randomly generated for this example be
\begin{equation*}
     B(x, u_1, u_2) = (8 + 7x) u_1 +(5+11x)u_2
\end{equation*}
and then product polynomials
\begin{align*} \label{toy-plain}
    P(x, u_1, u_2) &= f(x)B(x, u_1, u_2)  =(6+9x+11x^2)u_1 +(7+11x+8x^2)u_2 \bmod{13} \nonumber \\
    Q(x, u_1, u_2) &= h(x)B(x, u_1, u_2)  =(2+9x+10x^2)u_1 +(11+2x+12x^2)u_2 \bmod{13}
\end{align*}
Now let's encryption $P(x)$ and $Q(x)$ with $R_1$ over the rings
\begin{equation*} \label{toy-cipher}
\begin{aligned}
    \mathcal{P}(x, u_1, u_2) &= 4267* P(x, u_1, u_2) \bmod{6797}   \\
        &=(5211+4418x+6155x^2)u_1 +(2681+6155x+151x^2)u_2  \\
    \mathcal{Q}(x, u_1, u_2) &= 6475*Q(x, u_1, u_2) \bmod{7123} \\
                   &=(5827+1291x+643x^2)u_1 +(7118+5827x+6470x^2)u_2 
\end{aligned}
\end{equation*}
Then we obtain the key pair by choosing the Barrett parameter $R=2^{24}$:

\textbf{Private Key:}
\begin{align*}
    & f_0 = 4, f_1 = 9; h_0 =10, h_1=9 \\
    & R_1 = 4267, S_1=6797; R_2=6475, S_2 = 7123
\end{align*}

\textbf{Public Key:}
For simplification, we choose $\beta=1$ to evaluate all public key elements.
\begin{align*}
    & s_1 = S_1 \bmod{13} =11, s_2 = S_2 \bmod{13} = 12 \\
    & p'_{01} = P_{01} \bmod{13}= 5211 \bmod{13} = 11 \\
    & p'_{11} = P_{11} \bmod{13}= 4418 \bmod{13} = 11 \\
    & p'_{21} = P_{21} \bmod{13}= 6155 \bmod{13} = 6 \\
    & p'_{02} = P_{02} \bmod{13}= 2681 \bmod{13} = 3 \\
    & p'_{12} = P_{12} \bmod{13}= 6155 \bmod{13} = 6 \\
    & p'_{22} = P_{22} \bmod{13}= 151 \bmod{13} = 8 \\
    & q'_{01} = Q_{01} \bmod{13}= 5827 \bmod{13} = 3 \\
    & q'_{11} = Q_{11} \bmod{13}= 1291 \bmod{13} = 4 \\
    & q'_{21} = Q_{21} \bmod{13}= 643 \bmod{13} = 6 \\
    & q'_{02} = Q_{02} \bmod{13}= 7118 \bmod{13} = 7 \\
    & q'_{12} = Q_{12} \bmod{13}= 5827 \bmod{13} = 3 \\
    & q'_{22} = Q_{22} \bmod{13}= 6470 \bmod{13} = 9 \\
    & \mu'_{01} = \lfloor \frac{RP_{01}}{S_1} \rfloor = \lfloor \frac{2^24*5211}{6797} \rfloor = 12862449 \\
    & \mu'_{11} = \lfloor \frac{RP_{11}}{S_1} \rfloor = \lfloor \frac{2^24*4418}{6797} \rfloor = 10905066\\
    & \mu'_{21} =  \lfloor \frac{RP_{21}}{S_1} \rfloor = \lfloor \frac{2^24*6155}{6797} \rfloor = 15192550 \\
    & \mu'_{02} = \lfloor \frac{RP_{02}}{S_1} \rfloor = \lfloor \frac{2^24*2681}{6797} \rfloor = 6617583\\
    & \mu'_{12} = \lfloor \frac{RP_{12}}{S_1} \rfloor = \lfloor \frac{2^24*6155}{6797} \rfloor = 15192550\\
    & \mu'_{22} = \lfloor \frac{RP_{22}}{S_1} \rfloor = \lfloor \frac{2^24*151}{6797} \rfloor = 372717\\
    & \nu'_{01} = \lfloor \frac{RQ_{01}}{S_2} \rfloor = \lfloor \frac{2^24*5827}{7123} \rfloor = 13724671\\
    & \nu'_{11} = \lfloor \frac{RQ_{11}}{S_2} \rfloor = \lfloor \frac{2^24*1291}{7123} \rfloor = 3040767\\
    & \nu'_{21} = \lfloor \frac{RQ_{21}}{S_2} \rfloor = \lfloor \frac{2^24*643}{7123} \rfloor = 1514495\\
    & \nu'_{02} = \lfloor \frac{RQ_{02}}{S_2} \rfloor = \lfloor \frac{2^24*7118}{7123} \rfloor = 16765439\\
    & \nu'_{12} = \lfloor \frac{RQ_{12}}{S_2} \rfloor = \lfloor \frac{2^24*5827}{7123} \rfloor =  13724671\\
    & \nu'_{22} = \lfloor \frac{RQ_{22}}{S_2} \rfloor = \lfloor \frac{2^24*6470}{7123} \rfloor = 15239167\\
\end{align*}
The above public key together with security parameters $p=13, R=2^{24}$ would be available for any verifier to process any signature generated by the true signer holding the private key.

\subsubsection{Signing}
Assume that the hashed value of a message $M$ is $x=9$. Here are steps to generate the signature:
\begin{itemize}
    \item Choose a random $\alpha = 1 \in \mathbb{F}_{13}$,
    \item Evaluate $F= R_2^{-1} \times [\alpha f(x) \bmod{p}] \bmod{S_2} = 6475^{-1} \times [4+9*9 \bmod{13}] \bmod{7123}=5683 $
    \item Evaluate $H= R_1^{-1} \times [\alpha h(x) \bmod{p}] \bmod{S_1} = 4267^{-1} \times [10+ 7*9 \bmod{13}] \bmod{6797}=5357 $
\end{itemize}
then the signature is $sig=\{\textbf{5683}, \textbf{5357}\}.$
\subsubsection{Verify}
 After receiving the signature from the sender, the signature verification consists of two steps: evaluating the coefficients of verification polynomials $U(H, x, u_1, u_2)$ and  $V(F, x, u_1, u_2)$ with the signature elements and then polynomial values at the variable $x=Hash(M)$. Let's first prepare all coefficients:
 \begin{align*}
    & U_{01} = Hp'_{01}-s_1\lfloor \frac{H\mu_{01}}{R} \rfloor = 5357*11-11*\lfloor \frac{5357*12862449}{2^{24}} \rfloor \bmod{13}= 9 \\
    & U_{11} = Hp'_{11}-s_1\lfloor \frac{H\mu_{11}}{R} \rfloor = 5357*11-11*\lfloor \frac{5357*10905066}{2^{24}} \rfloor \bmod{13}=7\\
    & U_{21} =  Hp'_{21}-s_1\lfloor \frac{H\mu_{21}}{R} \rfloor = 5357*6-11*\lfloor \frac{5357*15192550}{2^{24}} \rfloor \bmod{13}=10 \\
    & U_{02} = Hp'_{02}-s_1\lfloor \frac{H\mu_{02}}{R} \rfloor = 5357*3-11*\lfloor \frac{5357*6617583}{2^{24}} \rfloor \bmod{13}=4\\
    & U_{12} = Hp'_{12}-s_1\lfloor \frac{H\mu_{12}}{R} \rfloor = 5357*6-11*\lfloor \frac{5357*1519255}{2^{24}} \rfloor \bmod{13}=10\\
    & U_{22} = Hp'_{22}-s_1\lfloor \frac{H\mu_{22}}{R} \rfloor = 5357*8-11*\lfloor \frac{5357*372717}{2^{24}} \rfloor \bmod{13}=12\\
    & V_{01} = Fq'_{01}-s_2\lfloor \frac{F\nu_{01}}{R} \rfloor = 5683*3-12*\lfloor \frac{5683*13724671}{2^{24}} \rfloor \bmod{13}=1\\
    & V_{11} = Fq'_{11}-s_2\lfloor \frac{F\nu_{11}}{R} \rfloor = 5683*4-12*\lfloor \frac{5683*3040767}{2^{24}} \rfloor \bmod{13}=11\\
    & V_{21} = Fq'_{21}-s_2\lfloor \frac{F\nu_{21}}{R} \rfloor = 5683*6-12*\lfloor \frac{5683*3040767}{2^{24}} \rfloor \bmod{13}=5\\
    & V_{02} = Fq'_{02}-s_2\lfloor \frac{F\nu_{02}}{R} \rfloor = 5683*7-12*\lfloor \frac{5683*16765439}{2^{24}} \rfloor \bmod{13}=12\\
    & V_{12} = Fq'_{12}-s_2\lfloor \frac{F\nu_{12}}{R} \rfloor = 5683*3-12*\lfloor \frac{5683*13724671}{2^{24}} \rfloor \bmod{13}=1\\
    & V_{22} = Fq'_{22}-s_2\lfloor \frac{F\nu_{22}}{R} \rfloor = 5683*9-12*\lfloor \frac{5683*15239167}{2^{24}} \rfloor \bmod{13}=6\\
\end{align*}
So we obtain the verification polynomials
\begin{equation}
\begin{aligned} \label{eq:toy-verify}
    &U(H=5357, x, u_1, u_2)= (9+7x+10x^2)u_1 +(4+10x+12x^2)u_2 \bmod{13} \\
    &V(F=5683, x, u_1, u_2)= (1+11x+5x^2)u_1 +(12+x+6x^2)u_2 \bmod{13}
\end{aligned}
\end{equation}
It is easy to verify that $U_1(x=9)=V_1(x=9)=11 \bmod{13}$ and $U_2(x=9)=V_2(x=9)=0 \bmod{13}$. For any randomly chosen $u_1, u_2 \in \mathbb{F}_{13}$, $U(F=5683, x=9, u_1, u_2)=V(H=5357, x=9, u_1, u_2)$. The verification is passed.

It is noticed that the signature of a given variable value $x$ would be randomized by choosing a random factor $\alpha \in \mathbb{F}_p$, which would lead to $''random''$ verification polynomials as shown in Eq.~\eqref{eq:toy-verify}. However, a genuine signature would pass the verification.

This toy example also demonstrates that $U_1(x)=V_1(x) \bmod{13} \longrightarrow 5x^2+9x+8= 5(x+3)(x+4)=0 \bmod{13}.$ There are two roots $r_1=9$ and $r_2=10$ satisfying the equation. That means, the same signature $\{F=5683, H=5357\}$ would pass the verification for $x=9$ and $x=10$, which leads a potential forged signature with $x=10$. Considering $U_2(x)=V_2(x) \bmod{13}\longrightarrow 6x^2+9x + 5 = 6(x+4)^2 =0 \bmod{13} $ which only has a single root $r_1=r_2=9 \in \mathbb{F}_{13}$. That means, $x=10$ would not satisfy the verification $U_2(x)=V_2(x) \bmod{13}$. Therefore, the minimum requirement is to set $m=2$ to avoid potential forged signature.  

\section{HPPK DS Security Analysis}\label{sec:security}
The MPPK DS scheme includes a verification equation system (Eq.~\eqref{eq:ds-identity1}) establishing a linear relationship between the public key and the coefficients of polynomials $P'(x, u_1, \dots, u_m)$, $Q'(x, u_1, \dots, u_m)$, and the private key, which consists of the coefficients of $f(x)$, $h(x)$, $s_0(x)$, and $s_n(x)$. However, the MPPK DS forgery attack~\cite{Guo2023} exploited the fact that any solution (not necessarily the true private key) for the private key from Eq.~\eqref{eq:ds-identity1} would pass the verification equation~\eqref{eq:ds-identity2}, leading to a forged signature.

On the other hand, the HPPK DS verification equation~\eqref{eq:verify} incorporates verification polynomials $U(x, u_1, \dots, u_m)$ and $V(x, u_1, \dots, u_m)$ with coefficients defined in Eq.~\eqref{eq:uv1}, effectively breaking the linearity between the private and public keys. Additionally, the Barrett reduction algorithm can be viewed as further encrypting the HPPK public key via a non-linear map $P_{ij}\rightarrow \mu_{ij}$, $Q_{ij}\rightarrow \nu_{ij}$, as shown in Eq.~\eqref{eq:public}. This non-linearity serves to thwart forgery attacks in the MPPK DS scheme.

In this section, we primarily outline the attacks we have identified on the HPPK DS scheme to date, along with their classical computational complexities. We focus on two primary avenues of attack: key recovery and signature forgery. Further detailed cryptanalysis will be conducted in future works.

\begin{proposition}
Assuming the same bit length $L$ for two hidden rings, there exists a private key recovery attack on the HPPK DS scheme with classical computational complexity of $\mathcal{O}(2^L)$, leveraging the HPPK DS public key and intercepted signatures.
\end{proposition}

\begin{proof}
Exploiting the public key elements $\mu_{ij}$ and $\nu_{ij}$ from Eq.~\eqref{eq:public}, we express $P_{ij}$ and $Q_{ij}$:

\begin{equation}
\begin{aligned}
&\mu_{ij} = \lfloor \frac{R \cdot P_{ij}}{S_1} \rfloor \longrightarrow P_{ij} = \lceil \frac{S_1 \cdot \mu_{ij}}{R} \rceil\ \\ 
&\nu_{ij} = \lfloor \frac{R \cdot Q_{ij}}{S_2} \rfloor \longrightarrow Q_{ij} = \lceil \frac{S_2 \cdot \nu_{ij}}{R} \rceil
\end{aligned}
\end{equation}
This refined attack involves iteratively searching for $S_1$ within the range $2^{L-1}$ to $2^L$, computing $P_{ij}$ using the known public key $\mu_{ij}$, and recalculating $p{''}_{ij}=P_{ij} \bmod{p}$ comparing with the public key $p'_{ij}$. Upon matching the recomputed $p{''}_{ij}$ with the public key $p'_{ij}$, the attacker deterministically identifies the private values $S_1$ and $P_{ij}$. The computational complexity of this approach is $\mathcal{O}(2^{L-1})$. A similar procedure applies to $S_2$ and $Q_{ij}$, resulting in a total complexity of $\mathcal{O}(2^L)$.

Moreover, the remaining private key elements can be readily determined by intercepting genuine signatures with the known values of $P_{ij}$, $Q_{ij}$:

\begin{equation}
\begin{aligned}
    &p_{ij} = H*P_{ij} \bmod{S_1} \bmod{p} = \alpha h(x) R_1^{-1} R_1 p_{ij} \bmod{S_1} \bmod{p}=\alpha h(x) B_{ij} \bmod{p}\\ 
    &q_{ij} = F*Q_{ij} \bmod{S_1} \bmod{p} = \alpha f(x) R_2^{-1} R_2 q_{ij} \bmod{S_2} \bmod{p}=\alpha f(x) B_{ij} \bmod{p}
\end{aligned}
\end{equation}

Then, $\frac{p_{ij}}{q_{ij}} = \frac{f(x)}{h(x)}$. By intercepting enough signatures, an attacker can find all coefficients of $f(x)$ and $h(x)$ except for a scalar factor which would not impact signature operations and pass the verification. Overall, this key recovery attack exhibits a complexity of $\mathcal{O}(2^{L})$.
\end{proof}


\begin{proposition}
Assuming the same bit length $L$ for two hidden rings, there exists a forgery attack on the HPPK DS scheme with classical computational complexity of $\mathcal{O}(2^{2L})$.
\end{proposition}

\begin{proof}
To spoof the signature, the adversary needs to find values $F',H'$ that satisfy 
\begin{equation}
    \sum_{k=0}^{\lambda + n}(F'q'_{kj}-s_2\lfloor \frac{F'\nu_{kj}}{R}\rfloor)x^k \mod p = \sum_{k=0}^{\lambda + n}(H'p'_{kj} - s_1\lfloor \frac{H'\mu_{kj}}{R}\rfloor)x^k \mod p,
\end{equation}
for all $j \in \{1, \dots, m\}$. Due to non-linearity of the floor functions in the above equation, an attacker requires to guess both elements of the signature $\{F', H'\} \in [1, 2^L)$ then evaluate if the above equation is true. If a set of $\{F', H'\}$ makes the above equation true, then a forgery signature is found and would pass the verification by a legal verifier. The overall complexity for this spoof attack is $\mathcal{O}(2^{2L})$, which is much worse than the private key recovery attack with $\mathcal{O}(2^{L})$. 
\end{proof}

In this section, we have briefly outlined  the attacks discovered to date on the HPPK DS algorithm. The most potent threats, from the attacker's perspective. The classical computational complexities of the attacks detailed in this section are denoted as $\mathcal{O}(2^L)$, with $L$ to be the bit length of hidden rings: $L=|S_1|_2 = |S_2|_2=2|p|_2+\delta$ with $\delta$ chosen based on the requirement of homomorphic property, culminating in the total classical computational complexity of the most effective attack on the HPPK DS algorithm being $\mathcal{O}(2^L).$ This, in essence, governs the sizing of the security parameters requisite for each security level. Then the key sizes and signature size can be directly calculated as follows:
\begin{itemize}
    \item \textbf{Public Key Size(B):} $2m(n+\lambda +1)(|R|_8 +|p|_8) +2|p|_8$.
    \item \textbf{Private Key Size(B):} $2(\lambda +1)|p|_8+4|L|_8$.
    \item \textbf{Signature Size(B):}  $2\frac{L*H}{|p|_8}$ with $H$ denoting the hash-code size of a selected hash algorithm required by NIST  and $|p|_8$ denoting the prime field size in bytes.
\end{itemize}

Due to the complexity independent from $n, \lambda, m$, we could optimally choose $n=\lambda=m=1$. 
Table~\ref{Tab:keyandsignsizes} presents a comprehensive overview of key sizes, signature sizes, and estimated entropy for all three NIST security levels. HPPK DS ensures robust security with entropy values of 144, 208, and 272 bits for NIST security levels I, III, and V, respectively.

The public key (PK) sizes exhibit a linear progression, with dimensions of 196, 276, and 356 bytes for security levels I, III, and V. Conversely, the private key (SK) sizes remain remarkably small, comprising 104 bytes for level I, 152 bytes for level III, and 200 bytes for level V.

In terms of signature sizes, HPPK DS maintains efficiency with compact outputs of 144, 208, and 272 bytes for NIST security levels I, III, and V, respectively. Our forthcoming work will delve into the benchmark performance of HPPK DS and provide a thorough comparison with NIST-standardized algorithms. 

\begin{table}[h!]
    \caption{The key and signature sizes in bytes, as provided by the HPPK DS scheme for the proposed parameter sets, are determined based on the optimal complexity of $\mathcal{O}(2^{L})$. In this context, we choose the Barrett parameter $R$ to be $32$ bits longer than $L$, and the hidden ring size is set to be $L=2\times|p|_2 + 16$ bits. All data is presented in bytes. The configuration is defined as $(log_2 p, n, \lambda, m, L, log_2R). $}
    \centering
\setlength{\tabcolsep}{10pt}
{\begin{tabular}{cccccccc} 
 \hline
Security & p& Configuration &Entropy (bits)  & $PK$& $SK$ & $Sig$ & $Hash$\\ 
 \hline
 I    &$2^{64}-59$ & (64,1,1,1, 144, 176)& 144& \textbf{196} & 104 & 144 & SHA-256 \\
 III  &$2^{96}-17$ &(96, 1,1,1, 208, 240)& 208&  \textbf{276} &152 & 208 & SHA-384\\
 V    &$2^{128}-159$&(128, 1,1,1, 272, 304)& 272&  \textbf{356} & 200& 272 & SHA-512\\
 \hline
\end{tabular}}
\label{Tab:keyandsignsizes}
\end{table}


\section{Conclusion}\label{conclusion}
To counteract the forgery attack observed in MPPK/DS~\cite{kuang2022.08.01-DS, kuang2023-ODS}, this paper suggests an extension of the HPPK KEM~\cite{kuang2023-HPPK-KEM} utilizing dual hidden rings for a quantum-secure digital signature scheme. The Barrett reduction algorithm for modular multiplication is expanded to convert modular multiplications over dual hidden rings into divisions by the Barrett parameter $R$. This embedding process incorporates signature elements into coefficients of a public polynomial over $\mathbb{F}_p$ through the floor function. The non-linear nature of the floor function contributes to the security of the proposed HPPK DS scheme. The functionality of the HPPK DS scheme is illustrated through a toy example. Security analysis reveals that the HPPK DS scheme achieves a complexity of $\mathcal{O}(2^L)$. Future work involves benchmarking its performance and comparing it with NIST-standardized algorithms.
\bibliography{my}



\section*{Acknowledgements}
 The authors extend their sincere gratitude to Michael Redding and Jay Toth for their continuous support and encouragement in tackling the challenges posed by forgery attacks in MPPK DS. Special thanks are also extended to Dr. Brian LaMacchia for engaging discussions throughout the evolution of HPPK DS developments. Additionally, we want to express our appreciation to one of the reviewers for bringing up the Merkle-Hellman knapsack cryptosystems, from which HPPK cryptography inherits its modular multiplication encryption. 

\section*{Author contributions statement}
R.K conceptualized the extension of HPPK KEM for the digital signature scheme and revised the manuscript based on reviewers' comments, M.P. conducted the security analysis, and M.S. and D.L. delved into the proposal and its security aspects. All authors contributed to and reviewed the manuscript.



The corresponding author is responsible for submitting a \href{http://www.nature.com/srep/policies/index.html#competing}{competing interests statement} on behalf of all authors of the paper. This statement must be included in the submitted article file.

\noindent \textbf{Data Availability ---}
All data generated or analyzed during this study is included in this published article.
\end{document}